\documentclass[%
 preprint,
preprintnumbers,
nofootinbib,
 amsmath,amssymb,
 aps,
]{revtex4-1}

\newcommand{\beq}{\begin{eqnarray}}
\newcommand{\eeq}{\end{eqnarray}}

\usepackage{graphicx}
\usepackage{dcolumn}
\usepackage{bm}
\usepackage{epstopdf}
\usepackage{wrapfig,epsfig}
\usepackage{mathrsfs}
\usepackage{color}
\usepackage[utf8]{inputenc} 

\newcommand{\Slash}[1]{{\ooalign{\hfil/\hfil\crcr$#1$}}}

\begin{document}
\preprint{J-PARC TH-0140}

\title{Three-loop formula for quark and gluon contributions\\ to the QCD trace anomaly}

\author{Kazuhiro Tanaka}
\email{kztanaka@juntendo.ac.jp}
\affiliation{Department of Physics, Juntendo University, Inzai, Chiba 270-1695, Japan }
\affiliation{J-PARC Branch, KEK Theory Center, Institute of Particle and Nuclear Studies, KEK, 203-1, Shirakata, Tokai, Ibaraki, 31901106, Japan}


\date{\today}

\begin{abstract}
In the QCD energy-momentum tensor $T^{\mu\nu}$,
the terms that contribute to physical matrix elements are expressed as the sum of 
the gauge-invariant quark part and gluon part.
Each part undergoes the renormalization due to the interactions among quarks and gluons,
although the total tensor $T^{\mu\nu}$ is not renormalized thanks to the conservation of energy and momentum.
Recently it has been shown that, through the renormalization, each of the quark and gluon parts of $T^{\mu\nu}$
receives
a definite amount of anomalous trace contribution, such that their sum reproduces the well-known QCD trace anomaly,
$T^\mu_\mu= (\beta/2g)F^{\mu\nu}F_{\mu\nu}+ m (1+\gamma_m)\bar{\psi}\psi$,
and the corresponding formulas have been derived up to two-loop order.
We extend this result to the three-loop order, working out all the relevant three-loop renormalization
structure for the quark and gluon energy-momentum tensors in the (modified) minimal subtraction scheme in the dimensional regularization.
We apply our three-loop formula of the quark/gluon decomposition of the trace anomaly 
to calculate the anomaly-induced mass structure of nucleons as well as pions.
\end{abstract}
\maketitle

\section{Introduction} 
\label{sec1}

The QCD energy-momentum tensor $T^{\mu\nu}$ is known to receive the trace anomaly as~\cite{Nielsen:1977sy,Adler:1976zt,Collins:1976yq}, 
\begin{equation}
T^\mu_\mu=\frac{\beta(g_R)}{2g_R}\left(F^{\mu\nu}F_{\mu\nu}\right)_R
+ \left(1+\gamma_m(g_R)\right)\left(m\bar{\psi}\psi\right)_R\ ,  
\label{111}
\end{equation}
representing the broken scale invariance due to the quantum loop effects,
with the beta-function $\beta$ for the QCD coupling constant $g$ 
and the anomalous dimension $\gamma_m$ for the quark mass $m$. 
Here, the suffix $R$ denotes the renormalized quantities, and
$\left(F^{\mu\nu}F_{\mu\nu}\right)_R$ and $\left(\bar{\psi}\psi\right)_R$ denote the renormalized composite operators.
This QCD trace anomaly signals the generation of a nonperturbative mass scale, say, the nucleon mass $M$,
taking the nucleon matrix element of (\ref{111}) and using the fact that $\langle P|T^{\mu\nu}|P\rangle= 2P^\mu P^\nu$, where $|P\rangle$ is the nucleon state with 4-momentum $P^\mu$.

The energy-momentum tensor consists of the quark part and the gluon part,  $T^{\mu\nu}=T_q^{\mu\nu} +T_g^{\mu\nu}$.
Considering the corresponding renormalized quantities as $\left(T^{\mu\nu}\right)_R=\left(T_q^{\mu\nu}\right)_R +\left(T_g^{\mu\nu}\right)_R$, we have 
\begin{equation}
\left(T^{\mu\nu}\right)_R=T^{\mu\nu}\ , \;\;\;\;\;\; \;\;\;\;\;\; \mbox{due to}\;\;\; \partial_\nu T^{\mu\nu}=0\ ,
\label{nonren}
\end{equation}
representing conservation of energy and momentum,
while $\left(T_q^{\mu\nu}\right)_R$ and $\left(T_g^{\mu\nu}\right)_R$ are not conserved separately and thus are different from 
$T_q^{\mu\nu}$ and $T_g^{\mu\nu}$, respectively,
by the ultraviolet (UV) subtraction terms to make the divergent bare quantities finite.
Such subtraction terms for the second rank symmetric tensors in various quantum field theories
appear to produce a finite contribution, when contracted with the metric tensor $\eta_{\mu\nu}$~\cite{Collins:1984xc}.
The explicit results for the individual quark and gluon parts in QCD
have been derived at the two-loop level in the minimal subtraction (MS) scheme in the dimensional regularization
in our recent paper~\cite{Hatta:2018sqd}, and the corresponding one-loop results read,
\begin{eqnarray}
\eta_{\mu\nu}\left(T_q^{\mu\nu}\right)_R &=& T_{q\;\; \mu}^{\ \mu}+
 \frac{\alpha_s}{4\pi} \left(\frac{n_f}{3}\left(F^2\right)_R+ \frac{4C_F}{3} \left(m\bar{\psi}\psi \right)_R \right)
\nonumber\\
&=&
 (m\bar{\psi}\psi)_R+ \frac{\alpha_s}{4\pi} \left(\frac{n_f}{3}(F^2)_R + \frac{4C_F}{3}(m\bar{\psi}\psi)_R\right)
\ , \label{050}
\\
\eta_{\mu\nu}\left(T_g^{\mu\nu}\right)_R &=& T_{g\;\; \mu}^{\ \mu}
-
\frac{\alpha_s}{4\pi} \left(\frac{n_f}{3}\left(F^2\right)_R+ \frac{4C_F}{3} \left(m\bar{\psi}\psi \right)_R \right)
\nonumber\\
&=&
 \frac{\alpha_s}{4\pi}\left(-\frac{11C_A}{6} (F^2)_R + \frac{14C_F}{3}(m\bar{\psi}\psi)_R\right)
\ , \label{051}
\end{eqnarray}
with
\begin{equation}
\alpha_s^R=\frac{g_R^2}{4\pi}\equiv \alpha_s\ ,
\end{equation}
such that their total sum, $\eta_{\mu\nu}\left(T_q^{\mu\nu}\right)_R+\eta_{\mu\nu}\left(T_g^{\mu\nu}\right)_R$,
reproduces the QCD trace anomaly~(\ref{111})
with the one-loop terms for the renormalization group (RG) functions,
\begin{eqnarray}
\frac{\beta(g_R)}{g_R}&=&-\beta_0 \frac{\alpha_s}{4\pi}+\cdots\ , \;\;\; \;\;\; \beta_0=\frac{11}{3}C_A-\frac{2n_f}{3}\ ,
\label{beta0}
\\
\gamma_m(g_R)&=& 
3C_F\frac{\alpha_s}{2\pi} +\cdots\ ,
\end{eqnarray}
being substituted, where $C_F=\frac{N_c^2-1}{2N_c}=\frac{4}{3}$ and $C_A=N_c=3$ for $N_c$ ($=3$) color.
Therefore, the quark-loop ($\propto \frac{2n_f}{3}$) and gluon-loop ($\propto - \frac{11}{3}C_A$)
contributions to the gluon self-energy (in the background field method, see, e.g., \cite{Peskin:1995ev}) directly determine the
$\left(F^2\right)_R$ terms in (\ref{050}) and (\ref{051}), respectively.  
Similar correspondence may be invoked to the relative weights, $4$ and $14$, of the $(m\bar{\psi}\psi)_R$ terms in (\ref{050}) and (\ref{051}) by interpreting that those weights correspond to the respective contributions 
caused by the quark and gluon internal propagators in
the quark self-energy.
Such simple correspondence between 
the loop contributions of the diagrams
and the decomposition of the total anomaly 
demonstrates that the decomposition as (\ref{050}) and (\ref{051}) is of physical origin and well-defined; 
this point is also supported by the fact~\cite{Hatta:2018sqd} that actually (\ref{050}) and (\ref{051}) are separately RG-invariant to one-loop order,
because $(m \bar{\psi}\psi)_R$ is an exactly RG-invariant quantity and $\alpha_s \left(F^2\right)_R$ is RG-invariant 
up to the corrections of order $\alpha_s^2$.
However, 
similar intuitive correspondence 
is not obvious in the two-loop results derived in \cite{Hatta:2018sqd}, and
each of quark/gluon parts exhibits the RG scale dependence. Therefore, 
the decomposition at two loops
is likely to depend
on regularization and renormalization schemes to handle the UV divergences.
Within the MS-like (MS,  $\overline{\rm MS}$) schemes in the dimensional regularization, their mutual relation is straightforward,  but 
the results using the other schemes are not known at present.

Physical relevance as well as 
phenomenological implications of such decomposition is also demonstrated in \cite{Hatta:2018sqd}, such that 
those RG properties of the quark/gluon parts of the trace anomaly allow us to constrain the twist-four gravitational form factor $\bar{C}_{q,g}$,
which arises as one of the gravitational form factors~\cite{Ji:1996ek,Kumano:2017lhr,Polyakov:2018zvc,Tanaka:2018wea,Polyakov:2018exb}
to parametrize hadron matirx elements of the quark/gluon parts of the QCD energy-momentum tensor,
$\langle P'|T^{\mu\nu}_{q,g}|P\rangle$.
In particular, the solution of the corresponding two-loop RG equations provides the model-independent
determination of the forward ($P'\to P$) value of $\bar{C}_{q,g}$, at the level of accuracy $\sim 10$\%.
Such quantitative constraint could have an impact on  
the descriptions of the shape deep inside the hadrons reflecting dynamics of quarks and gluons,
such as the pressure distributions inside the hadrons~\cite{Polyakov:2018guq,Polyakov:2018zvc,Teryaev:2016edw};
indeed, the recent results of the pressure distributions inside the nucleon~\cite{Burkert:2018bqq} are based on the determination of 
the 
gravitational form factors from
the behaviors 
of the generalized parton distributions (GPDs)~\cite{Polyakov:2002yz,Diehl:2003ny,Belitsky:2005qn},
which are 
obtained by experiments like deeply virtual Compton scattering (DVCS)~\cite{Mueller:1998fv,Ji:1996nm,Radyushkin:1996nd,Goeke:2001tz,Diehl:2003ny,Belitsky:2005qn}, 
deeply virtual meson production~\cite{Collins:1996fb,Goloskokov:2009ia}, meson-induced Drell-Yan 
production~\cite{Berger:2001zn,Goloskokov:2015zsa,Sawada:2016mao}, etc. (see also \cite{Shanahan:2018pib,Shanahan:2018nnv}).
As another phenomenological implication, 
the cross section of the near-threshold photoproduction of $J/\psi$ in $ep$ scattering~\cite{Joosten:2018gyo}
is sensitive to the $F^2$ part of the trace anomaly (\ref{111})~\cite{Kharzeev:1998bz},
which can be conveniently handled~\cite{Hatta:2018ina} through the $P'\to P$ behavior of 
the gravitational form factors that represent $\langle P'|\eta_{\mu\nu}\left(T_{q,g}^{\mu\nu}\right)_R|P\rangle$.  
Also, apparently, the decomposed quark and gluon contributions to the trace anomaly should provide a new insight
on understanding the origin of the nucleon mass, one of the main objectives of  the future Electron-Ion Collider.

The two-loop results corresponding to (\ref{050}) and (\ref{051})
have been derived~\cite{Hatta:2018sqd}
working out the renormalization mixing among the operators 
arising in the quark and gluon energy-momentum tensors,
where the two-loop anomalous dimensions for the second moment of the twist-two quark/gluon distribution functions,
as well as for the mixing of the twist-four operators $F^{\mu\nu}F_{\mu\nu}$, $m\bar{\psi}\psi$,
have been used as the necessary input informations.
Because all such input informations appear to be available at the three-loop order, 
we extend, in this paper, the quark/gluon decomposition of the QCD trace anomaly to the three-loop level.
We work out  
the renormalization-mixing structure relevant for the quark and gluon energy-momentum tensors
in the $\overline{\rm MS}$ scheme in Sec~\ref{sec2}, present their explicit results at three loops, and, as their direct consequence, 
derive the three-loop results of quark/gluon trace anomaly in Sec.~\ref{sec3}.
As an application of our results, the anomaly-induced mass structure of nucleon as well as pion is discussed in Sec.~\ref{sec4}.
Sec.~\ref{sec5} is reserved for conclusions.

\vspace{-0.5cm}
\section{Operator mixing in renormalization of  energy-momentum tensor}
\label{sec2}

We follow, here and in the following, the notations and conventions of \cite{Hatta:2018sqd}.
Our starting point is the gauge-invariant, symmetric QCD energy momentum tensor which is given by~\cite{Ji:1996ek} (see also the article by Jackiw in \cite{Treiman:1986ep})
\begin{equation}
T^{\mu\nu} 
=  T^{\mu\nu}_q+ T^{\mu\nu}_g \ ,
\label{tqg}
\end{equation}
where the quark and gluon parts read, respectively, in terms of the bare quark/gluon fields,
\begin{equation}
T^{\mu\nu}_q= i\bar{\psi}\gamma^{(\mu}\overleftrightarrow{D}^{\nu)}\psi\ ,
\;\;\;\;\;\;\;\;\;\;\;\;
T_g^{\mu\nu}= -F^{\mu\lambda}F^\nu_{\ \lambda} + \frac{\eta^{\mu\nu}}{4}F^2\ .
\label{tg}
\end{equation}
Here,
we have neglected the ghost and gauge fixing terms as they do not affect our final results. 
$T^{\mu\nu}$ is conserved and therefore it is a finite, scale-independent operator. However, $T_g^{\mu\nu}$ and $T_q^{\mu\nu}$ are not conserved separately (see (\ref{phys}), (\ref{eq2}) below) and are subject to regularization and renormalization. 
Their traceless part and trace part are expressed by the twist-two and twist-four operators, respectively.
Therefore, in order to derive the quark and gluon contributions to the quantum anomalies in their trace parts,
we have to work out the renormalization structure of the corresponding twist-two as well as twist-four operators
associated with (\ref{tg}), clarifying their renormalization mixing.
For this purpose, let us write
\begin{eqnarray}
&&O_1=-F^{\mu\lambda}F^{\nu}_{\ \lambda}\ , \label{o1}\\
&&O_2= \eta^{\mu\nu}F^2\ ,\\
&&O_3=i\bar{\psi}\gamma^{(\mu}\overleftrightarrow{D}^{\nu)} \psi\ , \label{o3}\\
&&O_4=\eta^{\mu\nu}m\bar{\psi}\psi\ ,
\end{eqnarray}
and, using this basis of operators $O_k$ ($k=1, 2, 3, 4$),  $T^{\mu\nu}$ of (\ref{tqg}) is expressed as
\begin{equation}
T^{\mu\nu}=O_1+\frac{O_2}{4}+O_3\ .
\end{equation}
We introduce the renormalized composite operators $O_k^R$ and the corresponding renormalization constants as
\begin{eqnarray}
O_1^R&=&Z_T O_1 + Z_M O_2 + Z_L O_3 + Z_S O_4\ , \label{o1ren}\\
O_2^R&=&Z_F O_2+Z_C O_4\ , \label{o2ren}\\
O_3^R&=&Z_\psi O_3 +Z_KO_4+ Z_Q O_1 +Z_B O_2\ ,  \label{pre} \\
O_4^R&=& O_4\ . \label{o4ren}
\end{eqnarray}
Here, for simplicity, we do not explicitly show the mixing with the equations-of-motion operators as well as the BRST-exact operators; their matrix elements sandwiched between physical states vanish (see e.g., \cite{Politzer:1980me,Collins:1984xc,Kodaira:1997ig,Kodaira:1998jn}),
so that they do not affect our final result.
We note that the composite operator $O_1$, as well as $O_3$, is a mixture of the twist-two and -four operators,
and the corresponding twist-four components can be expressed in terms of $O_2$ and $O_4$.
The formulas (\ref{o4ren}) and (\ref{o2ren}) reflect, respectively, that  $O_4$ is a RG-invariant operator,
and that the twist-four operator $O_2$ mixes with itself and another twist-four operator $O_4$.

The renormalization constants arising in (\ref{o1ren})-(\ref{pre}) can in principle be calculated by evaluating 
the Feynman diagrams for the loop corrections of the corresponding composite operators.
As we explain here, however, the renormalization constants can be determined without explicit evaluation
of the those loop diagrams, i.e., by utilizing certain anomalous dimensions in the literature which are available 
to the three-loop order.
First, subtracting the trace part from (\ref{o1}), (\ref{o3}), we obtain the traceless, twist-two part.
Denoting the corresponding renormalized and bare operators of twist-two as ($\ell=1, 3$)
\begin{eqnarray}
\widetilde{O}^R_\ell=O^R_\ell-{\rm traces}\ , \;\;\;\; \;\;\;\; \widetilde{O}_\ell=O_\ell-{\rm traces}\ , 
\label{trsubt}
\end{eqnarray} 
respectively, and subtracting those traces from both sides of (\ref{pre}) and (\ref{o1ren}),
we obtain,
\begin{eqnarray}
\widetilde O_3^R&=&Z_\psi \widetilde O_3 + Z_Q \widetilde O_1 \ ,  \label{pretl} \\
\widetilde O_1^R&=&Z_L \widetilde O_3 + Z_T \widetilde O_1 \ . \label{o1rentl}
\end{eqnarray}
The differentiation of these relations with respect to the renormalization scale $\mu$ yields
the RG equations of the twist-two, spin-2 quark and gluon operators,
which should coincide with the second moment of the Dokshitzer-Gribov-Lipatov-Altarelli-Parisi (DGLAP) 
evolution equations for the flavor-singlet part
of the unpolarized parton distribution functions:
\begin{equation}
  \frac{d}{d \ln \mu} \begin{pmatrix} \widetilde O_3^R(\mu)  \\ \widetilde O_1^R(\mu) \end{pmatrix}
=  - \begin{pmatrix}\widetilde \gamma_{qq}(\alpha_s) &\widetilde \gamma_{qg}(\alpha_s) \\
\widetilde \gamma_{gq}(\alpha_s) &\widetilde \gamma_{gg}(\alpha_s) \end{pmatrix}   \begin{pmatrix} \widetilde O_3^R(\mu) \\ \widetilde O_1^R(\mu)\end{pmatrix} ,  \label{one}
\end{equation}
with the anomalous dimension matrix,
\begin{equation}
\widetilde{\bm{\gamma}}(\alpha_s)\equiv \begin{pmatrix}\widetilde \gamma_{qq}(\alpha_s) &\widetilde \gamma_{qg}(\alpha_s) \\
\widetilde \gamma_{gq}(\alpha_s) &\widetilde \gamma_{gg}(\alpha_s) \end{pmatrix}  = \frac{\alpha_s}{4\pi} \begin{pmatrix} \frac{16}{3}C_F & -\frac{4n_f}{3} \\ -\frac{16}{3}C_F & \frac{4n_f}{3} \end{pmatrix}  +\cdots\ ,
\label{dglapk}
\end{equation}
where the ellipses denote the two- and higher-loop terms,
given as the second moment of the singlet
DGLAP kernel.
In mass-independent regularization schemes like the dimensional regularization,
the anomalous dimensions of (\ref{dglapk}) depend on the scale $\mu$ only through that of the strong coupling constant $\alpha_s(\mu)$. As a result,  (\ref{one}) is integrated to give
\begin{eqnarray}
 \begin{pmatrix} \widetilde O_3^R(\mu)  \\ \widetilde O_1^R(\mu) \end{pmatrix}
= {\rm T}_{(\alpha_s)} 
\exp\left( -\int_{\alpha_s(\mu')}^{\alpha_s(\mu)}d\alpha\frac{1}{\beta\left(\sqrt{4\pi \alpha}\ \right)}\sqrt{\frac{\pi}{\alpha}}\ 
\widetilde{\bm{\gamma}}(\alpha)\right)
   \begin{pmatrix} \widetilde O_3^R(\mu') \\ \widetilde O_1^R(\mu')\end{pmatrix} ,  \label{oneintegrate}
\end{eqnarray}
where the beta function is defined as
\begin{equation}
\beta (g_R) \equiv \frac{dg_R}{d\ln \mu}=\sqrt{\frac{\pi}{\alpha_s}}\ \frac{d\alpha_s}{d\ln \mu}\ ,
\label{betagR}
\end{equation}
and 
${\rm T}_{(\alpha_s)}$ means an ordering in the coupling constant such that the couplings increase from right to left
($\mu < \mu'$).
In particular, employing the dimensional regularization in $d=4-2\epsilon$ space-time dimensions, the $\mu' \to \infty$ limit of (\ref{oneintegrate})
leads to the result,
\begin{eqnarray}
 \begin{pmatrix} \widetilde O_3^R(\mu)  \\ \widetilde O_1^R(\mu) \end{pmatrix}
&=& {\rm T}_{(\alpha_s)} 
\exp\left( -\int_0^{\alpha_s(\mu)}d\alpha\frac{1}{\beta\left(\sqrt{4\pi \alpha}\ \right)}\sqrt{\frac{\pi}{\alpha}}\ 
\widetilde{\bm{\gamma}}(\alpha)\right)
   \begin{pmatrix} \widetilde O_3 \\ \widetilde O_1 \end{pmatrix} 
\nonumber\\
&\equiv& 
\begin{pmatrix} U_{qq}& U_{qg} \\
U_{gq} & U_{gg}
\end{pmatrix} 
\begin{pmatrix} \widetilde O_3 \\ \widetilde O_1 \end{pmatrix} ,  \label{oneintegrateinfty}
\end{eqnarray}
where (\ref{betagR}) implies,
\begin{equation}
\frac{\beta(g_R)}{g_R}=\frac{d\ln \alpha_s}{d\ln \mu^2}=-\epsilon -\sum_{n=0}^\infty\beta_n \left(\frac{\alpha_s}{4\pi}\right)^{n+1}\ .
\label{betaexact}
\end{equation}
This result (\ref{oneintegrateinfty}) determines the renormalization constants in (\ref{pretl}), (\ref{o1rentl})  
as
\begin{equation}
Z_\psi =U_{qq}\ ,\;\;\;\;\;\;\;\;Z_Q =U_{qg}\ ,\;\;\;\;\;\;\;\;Z_L=U_{gq}\ ,\;\;\;\;\;\;\;\;Z_T =U_{gg}\ ,
\label{soldglap}
\end{equation}
in terms of the anomalous dimensions of  (\ref{dglapk}) in the MS-like schemes to a desired accuracy.

The renormalization constants $Z_M$, $Z_S$ in (\ref{o1ren}),  
as well as $Z_K$, $Z_B$ in (\ref{pre}), are still to be determined. For this purpose,
we need to analyze the relevant renormalization structure at the twist-four level by treating   
the explicit form of the ``traces''  in (\ref{trsubt}). For the bare operators, it is 
straightforward to construct the corresponding contributions as
\begin{eqnarray}
\widetilde O_1&=&-F^{\mu\lambda}F^{\nu}_{\ \lambda}+\frac{\eta^{\mu\nu}}{d}F^2=O_1 + \frac{1}{d}O_2\ ,
\label{tracebare}\\
\widetilde O_3&=& i\bar{\psi}\gamma^{(\mu}\overleftrightarrow{D}^{\nu)} \psi- \frac{\eta^{\mu\nu}}{d}i\bar{\psi}\gamma^{(\lambda}\overleftrightarrow{D}_{\lambda)} \psi
=O_3 - \frac{1}{d}O_4
\ .
\end{eqnarray}
Here, in the second formula, we have used the equations of motion for the quark fields,
$\bar{\psi}\left( i\overleftarrow{\Slash{D}} +m\right)=\left(i\overrightarrow{\Slash{D}}-m\right)\psi=0$.
On the other hand, such straightforward manipulation to construct the trace terms is not useful 
for the renormalized operators in (\ref{trsubt}), because 
the trace operation and the renormalization do not commute,
i.e., $\eta_{\mu \nu}\left(F^{\mu\lambda}F^{\nu}_{\ \lambda}\right)_R \neq \left(F^{\mu\lambda}F_{\mu \lambda}\right)_R$,
$\eta_{\mu \nu}\left(i\bar{\psi}\gamma^{(\mu}\overleftrightarrow{D}^{\nu)} \psi\right)_R\neq \left(i\bar{\psi}\gamma^{(\lambda}\overleftrightarrow{D}_{\lambda)} \psi\right)_R$: 
this is a general phenomenon one encounters when 
renormalizing a certain operator $\theta^{\mu \nu}$, that behaves as a symmetric second rank tensor~\cite{Collins:1984xc,Suzuki:2013gza}.
In the dimensional regularization with $d=4-2\epsilon$ dimensions, the counter terms to subtract the UV divergences in the bare operator as,
\begin{equation}
\left(\theta^{\mu \nu}\right)_R =\theta^{\mu \nu}+ \left[ \mbox{counter terms}\right]\ , 
\end{equation}
may contain a term proportional
to $\eta^{\mu \nu}/\epsilon$. Then, the contribution of this type, $\eta^{\mu \nu}/\epsilon$,
produces, 
\begin{equation}
\eta_{\mu \nu}\frac{\eta^{\mu \nu}}{\epsilon}=\frac{4-2\epsilon}{\epsilon}=\frac{4}{\epsilon}-2\ ,
\end{equation}
for the trace of the renormalized operator, $\eta_{\mu \nu}\left(\theta^{\mu \nu}\right)_R$, in the MS,
while it produces $4/\epsilon$ as a counter term to define $\left(\theta^\mu_{\ \mu}\right)_R$,
giving rise to a finite difference between $\eta_{\mu \nu}\left(\theta^{\mu \nu}\right)_R$ and $\left(\theta^\mu_{\ \mu}\right)_R$.
Finite contributions of this type represent the trace anomaly in quantum field theories. In particular, such finite contributions 
for $\eta_{\mu \nu}\left(-F^{\mu\lambda}F^{\nu}_{\ \lambda}\right)_R$,  the trace of  $O_1^R$,  
appear to produce not only  
the contribution of $O_2^R$ but also of another twist-four operator $O_4^R$, 
by contract to (\ref{tracebare})~\cite{Hatta:2018sqd}; 
similarly, $\eta_{\mu \nu}\left(i\bar{\psi}\gamma^{(\mu}\overleftrightarrow{D}^{\nu)} \psi\right)_R$, the trace of  $O_3^R$,  
receives the contribution of  $O_4^R$ 
as well as $O_2^R$. Thus, we express the corresponding contributions as
\begin{eqnarray}
O^R_1&=&\widetilde{O}^R_1+\frac{-1+x_1}{d}O_2^R +\frac{y_1}{d} O_4^R\ ,
\label{traceren0}
\\
O^R_3&=&\widetilde{O}^R_3+\frac{x_3}{d}O_2^R +\frac{1+y_3}{d} O_4^R\ ,
\label{traceren}
\end{eqnarray}
using certain coefficients $x_1$, $y_1$, $x_3$, and $y_3$ of order $\alpha_s$ and higher, which are to be determined below.
Substituting (\ref{tracebare})-(\ref{traceren}) into (\ref{o1ren}) and (\ref{pre}), we obtain,
\begin{eqnarray}
\frac{-1+x_1}{d}O_2^R +\frac{y_1}{d} O_4^R
&  =&\left(-\frac{1}{d}Z_T+Z_M\right)O_2
+\left(\frac{1}{d}Z_L  +Z_S\right) O_4\ , 
\label{co1a} \\
\frac{x_3}{d}O_2^R +\frac{1+y_3}{d} O_4^R
&=& \left(\frac{1}{d}Z_\psi +Z_K\right) O_4
+\left(-\frac{1}{d}Z_Q+Z_B\right) O_2\ , \label{co2a}
\end{eqnarray}
where the twist-two operators dropped out using (\ref{pretl}), (\ref{o1rentl}),
and the further substitution of  (\ref{o2ren}) and (\ref{o4ren})
yields the relations among the twist-four bare operators as
\begin{eqnarray}
&&\left(\frac{1-x_1}{d}Z_F-\frac{1}{d}Z_T+Z_M\right)O_2
+\left(\frac{1-x_1}{d}Z_C-\frac{y_1}{d} +\frac{1}{d}Z_L  +Z_S\right) O_4=0\ , 
\label{co1ab} \\
&&\left(-\frac{x_3}{d}Z_F-\frac{1}{d}Z_Q+Z_B\right) O_2+  \left(-\frac{x_3}{d}Z_C-\frac{1+y_3}{d} +\frac{1}{d}Z_\psi +Z_K\right) O_4
=0\ , \label{co2ab}
\end{eqnarray}
leading to the four conditions,
\begin{eqnarray}
Z_M&=&\frac{1}{4-2\epsilon}Z_T+\frac{-1+x_1}{4-2\epsilon}Z_F\ ,
\label{cond1}\\
Z_S&=&-\frac{1}{4-2\epsilon}Z_L+ \frac{-1+x_1}{4-2\epsilon}Z_C+\frac{y_1}{4-2\epsilon}  \ , 
\label{cond2} \\
Z_B&=& \frac{1}{4-2\epsilon}Z_Q+ \frac{x_3}{4-2\epsilon}Z_F\ ,
\end{eqnarray}
\begin{eqnarray}
Z_K&=& -\frac{1}{4-2\epsilon}Z_\psi+  \frac{x_3}{4-2\epsilon}Z_C+ \frac{1+y_3}{4-2\epsilon} \ .
\label{cond4}
\end{eqnarray}
Recalling that $Z_T, Z_L, Z_Q$, and $Z_\psi$ are already determined as (\ref{soldglap}),
these conditions allow us to determine the renormalization constants in the LHS,
once $Z_F$, $Z_C$, and the coefficients $x_1, y_1, x_3, y_3$ are given.
We note that the Feynman diagram calculation of $Z_F$ and $Z_C$ is available to the two-loop order in the literature~\cite{Tarrach:1981bi}.

Now that explicit forms of all six renormalization constants arising in the RHS of (\ref{cond1})-(\ref{cond4}) are
available to a certain accuracy,
we are able to fix also the coefficients $x_1, y_1, x_3$, and $y_3$,  
invoking the following property of the 
renormalization constants in the MS-like schemes:
in this scheme, the renormalization constants take the form,
\begin{equation}
Z_X=\left(\delta_{X,T}+\delta_{X,\psi}+\delta_{X,F}\right)+\frac{a_X}{\epsilon}+\frac{b_X}{\epsilon^2}+\frac{c_X}{\epsilon^3}+\cdots\ ,
\label{zx}
\end{equation}
with $X=T,L,Q,\psi,F$, and $C$; here, $a_X, b_X, c_X, \ldots,$ are constants depending only on $\alpha_s$, and
$\delta_{X,X'}$ denotes the Kronecker 
symbol.
Laurent expansion 
of the RHS of (\ref{cond1})-(\ref{cond4})
about $\epsilon=0$ would produce the ${\cal O}(\epsilon^0)$ terms as
\begin{eqnarray}
&&\frac{1}{32}\left[ \left(8+4
   a_T+2 b_T+c_T+\cdots\right)+\left(-1+x_1\right) \left(8+ 4 a_F+2 b_F+c_F+\cdots \right)\right]\ ,
\\
&&\frac{1}{32}\left[ -\left(4
   a_L+2 b_L+c_L+\cdots\right)+\left(-1+x_1\right) \left( 4 a_C+2 b_C+c_C+\cdots \right)+8 y_1\right]\ ,
\\
&&\frac{1}{32}\left[ \left(4
   a_Q+2 b_Q+c_Q+\cdots\right)+x_3 \left(8+ 4 a_F+2 b_F+c_F+\cdots \right)\right]\ ,
\\
&&\frac{1}{32}\left[ -\left(8+4
   a_\psi+2 b_\psi+c_\psi+\cdots\right)+x_3 \left( 4 a_F+2 b_F+c_F+\cdots \right)+8(1+y_3)\right]\ ,
\label{msconds}
\end{eqnarray}
respectively, but
these four formulas equal zero because
$Z_M$, $Z_S$, $Z_B$, and $Z_K$
should also obey
the form of (\ref{zx}).
These four conditions allow us to determine $x_1, y_1, x_3$, and $y_3$ as power series in $\alpha_s$ 
to the accuracy same as the renormalization constants (\ref{zx}).
Then, this result of $x_1, y_1, x_3$, $y_3$
allows us 
to determine $Z_M$, $Z_S$, $Z_B$, and $Z_K$ to the same accuracy using (\ref{cond1})-(\ref{cond4});
furthermore, using (\ref{traceren0}) and (\ref{traceren}),  
the result of $x_1, y_1, x_3$, and $y_3$ allows us to derive the gluon/quark individual contributions to the trace anomaly 
as,
\begin{eqnarray}
\eta_{\mu\nu}\left(T^{\mu\nu}_g\right)_R&=&-\eta_{\mu \nu}\left(F^{\mu\lambda}F^{\nu}_{\ \lambda}\right)_R + \left(F^2\right)_R 
 \nonumber\\
 &=&x_1\left(F^2\right)_R +y_1\left(m\bar{\psi}\psi\right)_R\ ,
 \label{traceanog}
 \\
\eta_{\mu\nu}\left(T^{\mu\nu}_q\right)_R&=&\eta_{\mu\nu}\left(i\bar{\psi}\gamma^{(\mu}\overleftrightarrow{D}^{\nu)} \psi\right)_R
\nonumber\\
&=&\left(m\bar{\psi}\psi\right)_R+ x_3\left(F^2\right)_R + y_3 \left(m\bar{\psi}\psi\right)_R\ .
\label{traceanoq}
\end{eqnarray}
It is worth mentioning that the obtained results of the renormalization constants satisfy the following constraints,
\begin{eqnarray}
&&Z_T+Z_Q=1\ ,  \label{11} \\
&& Z_M+\frac{Z_F}{4}+Z_B=\frac{1}{4}\ ,\\
&& Z_L +Z_\psi= 1\ , \\
&& Z_S+\frac{Z_C}{4}+Z_K=0\ , \label{44}
\end{eqnarray}
such that $T^{\mu\nu}=\left( T^{\mu\nu}\right)_R$ of (\ref{nonren}) is obeyed
with (\ref{tqg})-(\ref{o4ren}). Using (\ref{co1a}), (\ref{co2a}), and (\ref{o2ren}), 
the above results (\ref{traceanog}) and (\ref{traceanoq}) for individual anomalies  
may be reexpressed as
\begin{eqnarray}
\eta_{\mu\nu}\left(T^{\mu\nu}_g\right)_R
 &=&\left(Z_F-Z_T+Z_Md\right)F^2
+\left(Z_C+ Z_L  +Z_Sd\right) m\bar{\psi}\psi
\ ,
 \label{traceanogre}
 \\
\eta_{\mu\nu}\left(T^{\mu\nu}_q\right)_R
&=&\left(-Z_Q+Z_Bd\right) F^2+ \left(Z_\psi +Z_Kd\right) m\bar{\psi}\psi
\ ,
\label{traceanoqre}
\end{eqnarray}
in terms of the bare operators multiplied by the renormalization constants;
adding these formulas and using the constraints (\ref{11})-(\ref{44}), we get
\begin{eqnarray}
\eta_{\mu\nu}\left(T^{\mu\nu}_q\right)_R+\eta_{\mu\nu}\left(T^{\mu\nu}_g\right)_R
 &=&\left[Z_F-1+\left(Z_M+Z_B\right)\left(4-2\epsilon\right)\right]F^2
\nonumber\\
&+&\left[Z_C+1 +\left(Z_S+Z_K\right)\left(4-2\epsilon\right)\right] m\bar{\psi}\psi
\nonumber\\
 &=&-\frac{\epsilon}{2} F^2+m\bar{\psi}\psi+\frac{\epsilon}{2} \left(Z_FF^2
+Z_Cm\bar{\psi}\psi\right)
\nonumber\\
&=&-\frac{\epsilon}{2} F^2+m\bar{\psi}\psi
\ .
\label{traceanoqgadd}
\end{eqnarray}
Here, in the last equality, we have used (\ref{o2ren}) and that
$\epsilon \left(F^2\right)_R\to 0$, as $\epsilon \to 0$.
Note that this final form coincides with the asymptotic ($\mu\to\infty$) limit of the RHS of (\ref{111})
with the use of (\ref{betaexact}), so that
\begin{eqnarray}
\eta_{\mu\nu}\left(T^{\mu\nu}_q\right)_R+\eta_{\mu\nu}\left(T^{\mu\nu}_g\right)_R&=&
\lim_{\mu\to\infty} \left[\frac{\beta(g_R(\mu))}{2g_R(\mu)}\left(F^2\right)_R
+ \left(1+\gamma_m(g_R(\mu))\right)\left(m\bar{\psi}\psi\right)_R \right]
\nonumber\\
&=&T^\lambda_\lambda
\ ,
\label{traceanoqgadd2}
\end{eqnarray}
where we have used the RG invariance of $T^\lambda_\lambda$.
This result shows that the consistency conditions (\ref{11})-(\ref{44})
ensure that the total sum of the individual anomaly contributions (\ref{traceanog}) and (\ref{traceanoq}) 
correctly reproduces the QCD trace anomaly (\ref{111}) at arbitrary order in $\alpha_s$.

The formulas derived above hold to arbitrary order in perturbation theory.
It is demonstrated in \cite{Hatta:2018sqd}
that the above logic leading to the formulas (\ref{traceanog}) and (\ref{traceanoq}) 
indeed works at the two-loop level:
substituting the anomalous dimension matrix $\widetilde{\bm{\gamma}}(\alpha_s)$ at two loops~\cite{Floratos:1981hs,Larin:1996wd} 
into (\ref{oneintegrateinfty}),
we are able to obtain the two-loop formulas of $Z_T, Z_L, Z_Q$, and $Z_\psi$ using (\ref{soldglap}), 
and, combined with the two-loop result~\cite{Tarrach:1981bi} of $Z_F$ and $Z_C$,
the coefficients $x_1, y_1, x_3$, and $y_3$ are uniquely determined to order $\alpha_s^2$.
As a result, the two-loop result for $Z_M$, $Z_S$, $Z_B$, $Z_K$ and that for (\ref{traceanog}) and (\ref{traceanoq})
are obtained, such that
their one-loop terms read (\ref{050}) and (\ref{051}). 

We emphasize that the input information necessary for this method
is the anomalous dimension matrix $\widetilde{\bm{\gamma}}(\alpha_s)$ of (\ref{dglapk}) and 
the renormalization constants $Z_F$ and $Z_C$ of (\ref{o2ren}).
The anomalous dimension matrix $\widetilde{\bm{\gamma}}(\alpha_s)$ for the flavor-singlet part of the
DGLAP evolution equation is now known to three-loop order~\cite{Larin:1996wd,Vogt:2004mw}.
On the other hand, for $Z_F$ and $Z_C$ beyond two loops, 
there seems no Feynman-diagram calculation performing the renormalization of the higher-twist scalar operators
$F^2$ and $\bar \psi \psi$.
Fortunately, however, the constraints imposed by the RG invariance of the energy-momentum tensor
are strong enough to determine the form of $Z_F$ as well as $Z_C$: using the RG invariance of
(\ref{traceanoqgadd}) and (\ref{111}), we have,
\begin{eqnarray}
-\frac{\epsilon}{2} F^2+m\bar{\psi}\psi&=&\frac{\beta(g_R)}{2g_R}\left(F^2\right)_R
+ \left(1+\gamma_m(g_R)\right)\left(m\bar{\psi}\psi\right)_R
\nonumber\\
&=&\frac{\beta(g_R)}{2g_R}Z_FF^2
+ \left(1+\gamma_m(g_R)+ \frac{\beta(g_R)}{2g_R}Z_C\right)m\bar{\psi}\psi\ ,
\end{eqnarray}
where we have substituted (\ref{o2ren}) and (\ref{o4ren}); this yields the relations,
\begin{eqnarray}
Z_F&=&-\epsilon\frac{g_R}{\beta(g_R)}=\frac{1}{1 +\displaystyle{\frac{1}{\epsilon}}\left[\beta_0 \frac{\alpha_s}{4\pi}+\beta_1 \left(\frac{\alpha_s}{4\pi}\right)^2+\beta_2 \left(\frac{\alpha_s}{4\pi}\right)^3+\cdots\right]}\ ,
\label{zf}\\
&&
\nonumber\\
Z_C&=&-2\gamma_m(g_R)\frac{g_R}{\beta(g_R)}=\frac{\displaystyle{\frac{2}{\epsilon}}\left[\gamma_{m0}\frac{\alpha_s}{4\pi} +  \gamma_{m1}\left(\frac{\alpha_s}{4\pi}\right)^2+\gamma_{m2}\left(\frac{\alpha_s}{4\pi}\right)^3+\cdots\right]}{1 +\displaystyle{\frac{1}{\epsilon}}\left[\beta_0 \frac{\alpha_s}{4\pi}+\beta_1 \left(\frac{\alpha_s}{4\pi}\right)^2+\beta_2 \left(\frac{\alpha_s}{4\pi}\right)^3+\cdots\right]}
\ ,
\label{zc}
\end{eqnarray}
where we have substituted (\ref{betaexact}) and 
the anomalous dimension for the quark mass in the form,
\begin{equation}
\gamma_m(g_R)= - \frac{d\ln  m_R}{d \ln \mu}=\gamma_{m0}\frac{\alpha_s}{4\pi} +  \gamma_{m1}\left(\frac{\alpha_s}{4\pi}\right)^2+\gamma_{m2}\left(\frac{\alpha_s}{4\pi}\right)^3+\cdots
\ .
\label{gammamg}
\end{equation}
The power series expansion of (\ref{zf}), (\ref{zc}) in $\alpha_s$ reads
\begin{eqnarray}
Z_F& =&   1
+ \frac{\alpha_s}{4\pi}\left( -\frac{\beta _0}{\epsilon }
\right)
+\left(\frac{\alpha_s}{4\pi}\right)^2\left[\frac{\beta _0^2}{\epsilon ^2}-\frac{\beta
   _1}{\epsilon }
\right]
%
%
+\left(\frac{\alpha_s}{4\pi}\right)^3\left[-\frac{\beta _0^3}{\epsilon ^3}+\frac{2 \beta _1 \beta
   _0}{\epsilon ^2}-\frac{\beta _2}{\epsilon }
\right]
+\cdots
\ ,
\label{zfrst}\\
Z_C& =&   
\frac{\alpha_s}{4\pi}\left( \frac{2 \gamma _{m0}}{\epsilon }
\right)
+\left(\frac{\alpha_s}{4\pi}\right)^2\left[\frac{2 \gamma
   _{m1}}{\epsilon }-\frac{2 \beta _0 \gamma _{m0}}{\epsilon
   ^2}
\right]
\nonumber\\
&+&\left(\frac{\alpha_s}{4\pi}\right)^3\left[\frac{2 \beta _0^2 \gamma _{m0}}{\epsilon ^3}-\frac{2 \left(\beta
   _1 \gamma _{m0}+\beta _0 \gamma _{m1}\right)}{\epsilon
   ^2}+\frac{2 \gamma _{m2}}{\epsilon }
\right]
+\cdots
\ .\label{zcrst}
\end{eqnarray}
These formulas completely determine $Z_F$ and $Z_C$ 
from the beta function $\beta$ and the mass anomalous dimension
$\gamma_m$ in the MS-like scheme.
Indeed, these formulas correctly reproduce the two-loop results of $Z_F$ and $Z_C$ of \cite{Tarrach:1981bi},
which were obtained by the Feynman diagram calculation.
As another check, 
substituting (\ref{zfrst}) and (\ref{zcrst}) into the $\mu$ derivative of (\ref{o2ren}),
\begin{eqnarray}
\frac{d O_2^R}{d\ln \mu}&=&\frac{dZ_F}{d\ln \mu} O_2+\frac{dZ_C}{d\ln \mu} O_4
=\frac{d\ln Z_F}{d\ln \mu}O_2^R +\left(\frac{d\ln Z_C}{d\ln \mu} -\frac{d\ln Z_F}{d\ln \mu}\right)Z_C O_4^R\ ,
\end{eqnarray}
the result can be recast into
the following RG equation, 
\begin{equation}
\frac{d}{d\ln \mu}\left(F^2\right)_R=-2\left(\frac{d\left[\beta(g_R)/2g_R\right]}{d\ln g_R}\left(F^2\right)_R
+\frac{d\gamma_m(g_R)}{d\ln g_R}\left(m\bar{\psi}\psi\right)_R\right)\ ,
\label{1111}
\end{equation}
which, alternatively, may be derived as a direct consequence of the fact that
the total trace anomaly (\ref{111}) is RG invariant, 
$\frac{d}{d\mu} T^\lambda_\lambda=0$.

The mass anomalous dimension (\ref{gammamg}) is now known to 
four-loop order in the literature~\cite{Chetyrkin:1997dh,Vermaseren:1997fq}.
Therefore, combined with the three-loop DGLAP anomalous dimensions for $\widetilde{\bm{\gamma}}(\alpha_s)$
in (\ref{one})-(\ref{soldglap}),
the formulas derived in this section allow us to
work out the new three-loop result for the gluon/quark trace anomalies (\ref{traceanog}) and (\ref{traceanoq}), 
which is presented in the next section.

\section{Results at three loops}
\label{sec3}

We present the three-loop results in the MS or $\overline{\rm MS}$ schemes, using the general formulas of Sec.~\ref{sec2}
with the three-loop input informations substituted.
For this, the beta-function (\ref{betaexact}) is given with $\beta_0$ of (\ref{beta0}) and
\begin{eqnarray}
\beta_1&=&\frac{34}{3}C_A^2-2C_Fn_f -\frac{10}{3}C_An_f\ ,
\label{beta1}\\
\beta_2&=&\frac{2857 C_A^3}{54}
-\frac{1}{2}
   n_f \left(\frac{1415 C_A^2}{27}+\frac{205 C_A C_F}{9}-2
   C_F^2\right)
+\frac{1}{4} n_f^2 \left(\frac{158 C_A}{27}+\frac{44 C_F}{9}\right)\ ,
\label{beta2}
\end{eqnarray}
and the mass anomalous dimension (\ref{gammamg}) reads~\cite{Chetyrkin:1997dh,Vermaseren:1997fq}
\begin{eqnarray}
 \gamma_{m0}&=&6C_F\ ,
\label{gammam0}\\
\gamma_{m1}&=& 
3C_F^2 + \frac{97}{3}C_F C_A -\frac{10}{3}C_F n_f\ ,
\\
\gamma_{m2}&=& 
n_f \left[\left(-48 \zeta (3)-\frac{556}{27}\right) C_A C_F+(48 \zeta (3)-46)
   C_F^2\right]
\nonumber\\ &&
-\frac{129}{2} C_A C_F^2 
+\frac{11413}{54} C_A^2 C_F-\frac{70}{27}
   C_F n_f^2+129 C_F^3
\ ,
\label{gammam2}
\end{eqnarray}
where $\zeta(s)$ is the Riemann zeta-function with $\zeta(3)= 1.202056903\ldots$.
The three-loop anomalous dimension matrix (\ref{dglapk}) for the twist-two flavor-singlet operators reads~\cite{Larin:1996wd,Vogt:2004mw}
\begin{equation}
\widetilde{\bm{\gamma}}(\alpha_s)
= \frac{\alpha_s}{4\pi}\begin{pmatrix} \frac{16C_F}{3}\;\; &- \frac{4n_f}{3} \\ -\frac{16C_F}{3} \;\; & \frac{4n_f}{3} \end{pmatrix}+ \left(\frac{\alpha_s}{4\pi}\right)^2 \bm{S} +\left(\frac{\alpha_s}{4\pi}\right)^3  \bm{R}\ ,
\label{dglap}
\end{equation}
where 
\begin{eqnarray}
\bm{S}&=&2\begin{pmatrix} \frac{376}{27}C_FC_A  -\frac{112}{27} C_F^2 -\frac{104}{27} n_f C_F \;\;\;\;\;\;\; & 
-\frac{74}{27}C_Fn_f -\frac{35}{27}C_An_f \\
-\frac{376}{27}C_F C_A + \frac{112}{27}C_F^2 + \frac{104}{27}C_F n_f \;\;\;\;\; \;\;& \frac{74}{27}C_F n_f + \frac{35}{27}C_A n_f  \end{pmatrix}\ ,
\end{eqnarray}  
and
\begin{eqnarray}
R_{qq}=-R_{gq}&=&-\frac{256}{3} \zeta (3) C_A C_F n_f-\frac{44}{9} C_A C_F n_f-128 \zeta (3) C_A
   C_F^2+\frac{128}{3} \zeta (3) C_A^2 C_F
\nonumber\\&&
-\frac{17056}{243} C_A
   C_F^2
+\frac{41840}{243} C_A^2 C_F+\frac{256}{3} \zeta (3) C_F^2
   n_f-\frac{14188}{243} C_F^2 n_f
\nonumber\\&&
-\frac{568}{81} C_F n_f^2+\frac{256 \zeta (3)
   C_F^3}{3}-\frac{1120 C_F^3}{243}\ ,
\\
R_{qg}=-R_{gg}&=&
-\frac{208}{3} \zeta (3) C_A C_F n_f+\frac{278}{9} C_A C_F n_f+48 \zeta (3) C_A^2
   n_f-\frac{3589}{81} C_A^2 n_f
\nonumber\\&&
+\frac{2116}{243} C_A n_f^2+\frac{64}{3} \zeta
   (3) C_F^2 n_f-\frac{346}{243} C_F n_f^2-\frac{4310}{243} C_F^2 n_f
\ .
\end{eqnarray}
Then, using (\ref{soldglap}), we obtain the corresponding three-loop renormalization constants
as 
\begin{eqnarray}
Z_\psi   &=&  
1+ \frac{\alpha_s}{4\pi}\left( 
\frac{8 C_F}{3 \epsilon }
\right)
+\left(\frac{\alpha_s}{4\pi}\right)^2\left[
\frac{C_F \left(\frac{16 n_f}{9}-\frac{44 C_A}{9}\right)+\frac{32
   C_F^2}{9}}{\epsilon ^2}+\frac{C_F \left(\frac{188 C_A}{27}-\frac{52 n_f}{27}\right)-\frac{56
   C_F^2}{27}}{\epsilon }
\right]
\nonumber\\&&
+\left(\frac{\alpha_s}{4\pi}\right)^3\Bigl[
\frac{1}{\epsilon^3}\left\{
n_f \left(\frac{320 C_F^2}{81}-\frac{616 C_A
   C_F}{81}\right)-\frac{352}{27} C_A C_F^2+\frac{968}{81} C_A^2
   C_F
\right. \nonumber\\ &&\left.
+\frac{32}{27} C_F n_f^2+\frac{256 C_F^3}{81}
\right\}
+\frac{1}{\epsilon^2}\left\{
n_f \left(\frac{3272 C_A
   C_F}{243}-\frac{560 C_F^2}{243}\right)+\frac{5744}{243} C_A
   C_F^2
\right. \nonumber\\ &&\left.
-\frac{6584}{243} C_A^2 C_F-\frac{104}{81} C_F n_f^2-\frac{448
   C_F^3}{81}
\right\}
\nonumber\\&&
+\frac{1}{\epsilon}\left\{
n_f \left(-\frac{2}{27} (192 \zeta (3)+11) C_A C_F-\frac{2}{729}
   (3547-5184 \zeta (3)) C_F^2\right)
\right. \nonumber\\ &&\left.
-\frac{16}{729} (972 \zeta (3)+533) C_A
   C_F^2+\frac{8}{729} (648 \zeta (3)+2615) C_A^2 C_F
\right. \nonumber
\\ 
&&
\left.
-\frac{284}{243} C_F
   n_f^2+\frac{16}{729} (648 \zeta (3)-35) C_F^3
\right\}
\Bigr]
\ ,
\\
 Z_Q & =&   
 \frac{\alpha_s}{4\pi}\left( 
-\frac{2 n_f}{3 \epsilon }
\right)
+\left(\frac{\alpha_s}{4\pi}\right)^2\left[
\frac{\frac{11 C_A n_f}{9}-\frac{8 C_F n_f}{9}-\frac{4
   n_f^2}{9}}{\epsilon ^2}+\frac{-\frac{35}{54} C_A n_f-\frac{37 C_F n_f}{27}}{\epsilon }
\right]
\nonumber\\&&
+\left(\frac{\alpha_s}{4\pi}\right)^3\Bigl[
\frac{1}{\epsilon^3}\left\{
n_f^2 \left(\frac{154 C_A}{81}-\frac{80 C_F}{81}\right)+n_f \left(\frac{88
   C_A C_F}{27}-\frac{242 C_A^2}{81}-\frac{64 C_F^2}{81}\right)-\frac{8
   n_f^3}{27}
\right\}
\nonumber\\&&
+\frac{1}{\epsilon^2}\left\{
n_f^2 \left(-\frac{355 C_A}{243}-\frac{10 C_F}{9}\right)+n_f
   \left(-\frac{26 C_A C_F}{81}+\frac{997 C_A^2}{243}-\frac{8
   C_F^2}{27}\right)
\right\}
\nonumber
\\
&&
+\frac{1}{\epsilon}\left\{
n_f \left(\frac{1}{81} (417-936 \zeta (3)) C_A
   C_F+\frac{1}{486} (3888 \zeta (3)-3589) C_A^2
\right. \right. \nonumber
\\ 
&&\left.\left.
+\frac{1}{729} (2592 \zeta
   (3)-2155) C_F^2\right)+n_f^2 \left(\frac{1058 C_A}{729}-\frac{173
   C_F}{729}\right)
\right\}
\Bigr]
\ ,\label{zqthree}
\end{eqnarray}
\begin{eqnarray}
Z_L& =&   
 \frac{\alpha_s}{4\pi}\left( 
-\frac{8 C_F}{3 \epsilon }
\right)
+\left(\frac{\alpha_s}{4\pi}\right)^2\left[
\frac{C_F \left(\frac{44 C_A}{9}-\frac{16
   n_f}{9}\right)-\frac{32 C_F^2}{9}}{\epsilon ^2}+\frac{C_F \left(\frac{52 n_f}{27}-\frac{188
   C_A}{27}\right)+\frac{56 C_F^2}{27}}{\epsilon }
\right]
\nonumber\\&&
+\left(\frac{\alpha_s}{4\pi}\right)^3\Bigl[
\frac{1}{\epsilon^3}\left\{
n_f \left(\frac{616 C_A C_F}{81}-\frac{320
   C_F^2}{81}\right)+\frac{352}{27} C_A C_F^2-\frac{968}{81} C_A^2
   C_F
\right. \nonumber\\ &&\left.
-\frac{32}{27} C_F n_f^2
-\frac{256 C_F^3}{81}
\right\}
+\frac{1}{\epsilon^2}\left\{
n_f \left(\frac{560
   C_F^2}{243}-\frac{3272 C_A C_F}{243}\right)
\right. \nonumber\\ &&\left.
-\frac{5744}{243} C_A
   C_F^2+\frac{6584}{243} C_A^2 C_F
+\frac{104}{81} C_F n_f^2+\frac{448
   C_F^3}{81}
\right\}
\nonumber\\&&
+\frac{1}{\epsilon}\left\{
n_f \left(\frac{2}{27} (192 \zeta (3)+11) C_A C_F+\frac{2}{729}
   (3547-5184 \zeta (3)) C_F^2\right)
\right. \nonumber\\ &&\left.
+\frac{16}{729} (972 \zeta (3)+533) C_A
   C_F^2-\frac{8}{729} (648 \zeta (3)+2615) C_A^2 C_F
\right. \nonumber\\ &&\left.
+\frac{284}{243} C_F
   n_f^2-\frac{16}{729} (648 \zeta (3)-35) C_F^3
\right\}
\Bigr]
\ ,
\\
Z_T& =&   
1
+ \frac{\alpha_s}{4\pi}\left( 
\frac{2 n_f}{3 \epsilon }
\right)
+\left(\frac{\alpha_s}{4\pi}\right)^2\left[
\frac{-\frac{11 C_A n_f}{9}+\frac{8 C_F n_f}{9}+\frac{4
   n_f^2}{9}}{\epsilon ^2}+\frac{\frac{35 C_A n_f}{54}+\frac{37 C_F n_f}{27}}{\epsilon }
\right]
\nonumber\\&&
+\left(\frac{\alpha_s}{4\pi}\right)^3\Bigl[
\frac{1}{\epsilon^3}\left\{
n_f^2 \left(\frac{80 C_F}{81}-\frac{154 C_A}{81}\right)+n_f
   \left(-\frac{88 C_A C_F}{27}+\frac{242 C_A^2}{81}+\frac{64
   C_F^2}{81}\right)+\frac{8 n_f^3}{27}
\right\}
\nonumber\\&&
+\frac{1}{\epsilon^2}\left\{
n_f^2 \left(\frac{355 C_A}{243}+\frac{10
   C_F}{9}\right)+n_f \left(\frac{26 C_A C_F}{81}-\frac{997 C_A^2}{243}+\frac{8
   C_F^2}{27}\right)
\right\}
\nonumber\\&&
+\frac{1}{\epsilon}\left\{
n_f \left(\frac{1}{81} (936 \zeta (3)-417) C_A
   C_F+\frac{1}{486} (3589-3888 \zeta (3)) C_A^2
\right. \right.\nonumber\\ &&\left.\left.
+\frac{1}{729} (2155-2592 \zeta
   (3)) C_F^2\right)+n_f^2 \left(\frac{173 C_F}{729}-\frac{1058
   C_A}{729}\right)
\right\}
\Bigr]
\ .
\end{eqnarray}
On the other hand,
$Z_F$ and $Z_C$ at three loops are given by (\ref{zfrst}) and (\ref{zcrst}), respectively,
with (\ref{beta0}), (\ref{beta1})-(\ref{gammam2}) substituted.

Substituting the above formulas into (\ref{cond1})-(\ref{msconds}), 
the three-loop result for the remaining renormalization constants is obtained as
\begin{eqnarray}
Z_M & =&   
 \frac{\alpha_s}{4\pi}\left( 
\frac{11 C_A}{12 \epsilon }
\right)
+\left(\frac{\alpha_s}{4\pi}\right)^2\left[
\frac{\frac{11 C_A n_f}{12}-\frac{121 C_A^2}{36}+\frac{2 C_F
   n_f}{9}}{\epsilon ^2}+\frac{-\frac{14 C_A n_f}{27}+\frac{17 C_A^2}{6}-\frac{5 C_F
   n_f}{108}}{\epsilon }
\right]
\nonumber\\&&
+\left(\frac{\alpha_s}{4\pi}\right)^3\Bigl[
\frac{1}{\epsilon^3}\left\{
n_f^2 \left(\frac{121 C_A}{162}+\frac{20 C_F}{81}\right)+n_f
   \left(-\frac{22 C_A C_F}{27}-\frac{484 C_A^2}{81}+\frac{16
   C_F^2}{81}\right)
\right. \nonumber\\ &&\left.
+\frac{1331 C_A^3}{108}
\right\}
+\frac{1}{\epsilon^2}\left\{
n_f^2 \left(-\frac{140
   C_A}{243}-\frac{43 C_F}{162}\right)+n_f \left(\frac{541 C_A
   C_F}{162}+\frac{7889 C_A^2}{972}+\frac{14 C_F^2}{81}\right)
\right. \nonumber
\\
  &&\left.
-\frac{187
   C_A^3}{9}
\right\}
+\frac{1}{\epsilon}\left\{
n_f \left(\left(\frac{26 \zeta (3)}{9}-\frac{1123}{324}\right) C_A
   C_F+\left(-2 \zeta (3)-\frac{293}{72}\right)
   C_A^2
\right.\right. \nonumber
\\
  &&\left.\left.
+\left(\frac{784}{729}-\frac{8 \zeta (3)}{9}\right) C_F^2\right)+n_f^2
   \left(\frac{361 C_F}{1458}-\frac{655 C_A}{5832}\right)+\frac{2857
   C_A^3}{216}
\right\}
\Bigr]
\ ,
\end{eqnarray}
\begin{eqnarray}
Z_B& =&   
 \frac{\alpha_s}{4\pi}\left( 
-\frac{n_f}{6 \epsilon }
\right)
+\left(\frac{\alpha_s}{4\pi}\right)^2\left[
\frac{\frac{11 C_A n_f}{36}-\frac{2 C_F
   n_f}{9}-\frac{n_f^2}{9}}{\epsilon ^2}+\frac{-\frac{17}{54} C_A n_f-\frac{49 C_F
   n_f}{108}}{\epsilon }
\right]
\nonumber\\&&
+\left(\frac{\alpha_s}{4\pi}\right)^3\Bigl[
\frac{1}{\epsilon^3}\left\{
n_f^2 \left(\frac{77 C_A}{162}-\frac{20 C_F}{81}\right)+n_f \left(\frac{22
   C_A C_F}{27}-\frac{121 C_A^2}{162}-\frac{16 C_F^2}{81}\right)-\frac{2
   n_f^3}{27}
\right\}
\nonumber\\&&
+\frac{1}{\epsilon^2}\left\{
n_f^2 \left(-\frac{130 C_A}{243}-\frac{65 C_F}{162}\right)+n_f
   \left(\frac{53 C_A C_F}{162}+\frac{1723 C_A^2}{972}-\frac{14
   C_F^2}{81}\right)
\right\}
\nonumber\\&&
+\frac{1}{\epsilon}\left\{
n_f \left(\left(\frac{401}{648}-\frac{26 \zeta
   (3)}{9}\right) C_A C_F+\left(2 \zeta (3)-\frac{67}{27}\right)
   C_A^2+\left(\frac{8 \zeta (3)}{9}-\frac{2407}{2916}\right) C_F^2\right)
\right. \nonumber\\ &&\left.
+n_f^2
   \left(\frac{697 C_A}{1458}+\frac{169 C_F}{2916}\right)
\right\}
\Bigr]
\ ,
\label{zbthree}\\
Z_S &=&
 \frac{\alpha_s}{4\pi}\left( 
-\frac{7 C_F}{3 \epsilon }
\right)
+\left(\frac{\alpha_s}{4\pi}\right)^2\left[
\frac{C_F \left(\frac{88 C_A}{9}-\frac{14 n_f}{9}\right)+\frac{8
   C_F^2}{9}}{\epsilon ^2}+\frac{C_F \left(\frac{11 n_f}{27}-\frac{406 C_A}{27}\right)-\frac{85
   C_F^2}{54}}{\epsilon }
\right]
\nonumber\\&&
+\left(\frac{\alpha_s}{4\pi}\right)^3\Bigl[
\frac{1}{\epsilon^3}\left\{
n_f \left(\frac{1034 C_A C_F}{81}+\frac{80 C_F^2}{81}\right)-\frac{88}{27}
   C_A C_F^2-\frac{3025}{81} C_A^2 C_F
\right. \nonumber\\ &&\left.
-\frac{28}{27} C_F n_f^2+\frac{64
   C_F^3}{81}
\right\}
+\frac{1}{\epsilon^2}\left\{
n_f \left(-\frac{5056 C_A C_F}{243}-\frac{1721
   C_F^2}{243}\right)+\frac{4753}{486} C_A C_F^2
\right. \nonumber\\ &&\left.
+\frac{42767}{486} C_A^2
   C_F+\frac{22}{81} C_F n_f^2-\frac{80 C_F^3}{81}
\right\}
+\frac{1}{\epsilon}\left\{
n_f \left(\left(\frac{184
   \zeta (3)}{9}+\frac{1423}{243}\right) C_A
   C_F
\right. \right.\nonumber\\ &&\left.\left.
+\left(\frac{25229}{1458}-\frac{184 \zeta (3)}{9}\right)
   C_F^2\right)+\left(\frac{91753}{2916}-\frac{16 \zeta (3)}{3}\right) C_A
   C_F^2
\right. \nonumber\\ &&\left.
+\left(\frac{16 \zeta (3)}{9}-\frac{294929}{2916}\right) C_A^2
   C_F+\frac{277}{243} C_F n_f^2+\left(\frac{32 \zeta
   (3)}{9}-\frac{95041}{1458}\right) C_F^3
\right\}
\Bigr]
\ ,
\\
Z_K& =&   
 \frac{\alpha_s}{4\pi}\left( 
-\frac{2 C_F}{3 \epsilon }
\right)
+\left(\frac{\alpha_s}{4\pi}\right)^2\left[
\frac{C_F \left(\frac{11 C_A}{9}-\frac{4 n_f}{9}\right)-\frac{8
   C_F^2}{9}}{\epsilon ^2}+\frac{C_F \left(\frac{34 n_f}{27}-\frac{61 C_A}{54}\right)+\frac{2
   C_F^2}{27}}{\epsilon }
\right]
\nonumber\\&&
+\left(\frac{\alpha_s}{4\pi}\right)^3\Bigl[
\frac{1}{\epsilon^3}\left\{
n_f \left(\frac{154 C_A C_F}{81}-\frac{80 C_F^2}{81}\right)+\frac{88}{27}
   C_A C_F^2-\frac{242}{81} C_A^2 C_F
\right. \nonumber\\ &&\left.
-\frac{8}{27} C_F n_f^2-\frac{64
   C_F^3}{81}
\right\}
+\frac{1}{\epsilon^2}\left\{
n_f \left(\frac{20 C_F^2}{243}-\frac{1478 C_A
   C_F}{243}\right)-\frac{1040}{243} C_A C_F^2
\right. \nonumber\\ &&\left.
+\frac{1283}{243} C_A^2
   C_F+\frac{68}{81} C_F n_f^2+\frac{80 C_F^3}{81}
\right\}
+\frac{1}{\epsilon}\left\{
n_f \left(\left(\frac{32 \zeta
   (3)}{9}+\frac{1079}{243}\right) C_A C_F
\right.\right. \nonumber
\\ 
&&\left.\left.
+\left(\frac{8305}{1458}-\frac{32 \zeta
   (3)}{9}\right) C_F^2\right)
+\left(\frac{16 \zeta
   (3)}{3}+\frac{572}{729}\right) C_A C_F^2
\right. \nonumber
\\
&&\left.
+\left(-\frac{16 \zeta
   (3)}{9}-\frac{6611}{1458}\right) C_A^2 C_F
+\frac{38}{243} C_F
   n_f^2+\left(\frac{500}{729}-\frac{32 \zeta (3)}{9}\right) C_F^3
\right\}
\Bigr]
\ ,
\end{eqnarray}
and the coefficients in (\ref{traceren0}) and (\ref{traceren}) for the trace after renormalization are obtained as  
\begin{eqnarray}
x_1& =&   
 \frac{\alpha_s}{4\pi}\left( 
-\frac{11 C_A}{6}
\right)
+\left(\frac{\alpha_s}{4\pi}\right)^2\left[
n_f \left(\frac{28 C_A}{27}+\frac{5
   C_F}{54}\right)-\frac{17 C_A^2}{3}
\right]
%
%
\nonumber
\end{eqnarray}
\begin{eqnarray}
&&
+\left(\frac{\alpha_s}{4\pi}\right)^3\left[
n_f \left(\left(\frac{1123}{162}-\frac{52
   \zeta (3)}{9}\right) C_A C_F+\left(4 \zeta (3)+\frac{293}{36}\right)
   C_A^2
 \right.  \right. \nonumber\\ &&\left.\left.
+\frac{16}{729} (81 \zeta (3)-98) C_F^2\right)
+n_f^2 \left(\frac{655
   C_A}{2916}-\frac{361 C_F}{729}\right)-\frac{2857 C_A^3}{108}
\right]
\ ,
\\
y_1& =&   
 \frac{\alpha_s}{4\pi}\left( 
\frac{14 C_F}{3}
\right)
+\left(\frac{\alpha_s}{4\pi}\right)^2\left[
\frac{812 C_A C_F}{27}-\frac{22 C_F n_f}{27}+\frac{85
   C_F^2}{27}
\right]
%
%
\nonumber\\&&
+\left(\frac{\alpha_s}{4\pi}\right)^3\left[
n_f \left(\left(\frac{368 \zeta (3)}{9}-\frac{25229}{729}\right)
   C_F^2-\frac{2}{243} (4968 \zeta (3)+1423) C_A C_F\right)
 \right. \nonumber\\ &&\left.
+\left(\frac{32 \zeta
   (3)}{3}-\frac{91753}{1458}\right) C_A C_F^2+\left(\frac{294929}{1458}-\frac{32
   \zeta (3)}{9}\right) C_A^2 C_F-\frac{554}{243} C_F
   n_f^2
 \right. \nonumber\\ &&\left.
+\left(\frac{95041}{729}-\frac{64 \zeta (3)}{9}\right) C_F^3
\right]
\ ,
\\
x_3
& =&    \frac{\alpha_s}{4\pi}\left( 
\frac{n_f}{3}
\right)
+\left(\frac{\alpha_s}{4\pi}\right)^2\left[
\frac{17 C_A n_f}{27}+\frac{49 C_F n_f}{54}
\right]
\nonumber\\&&
+\left(\frac{\alpha_s}{4\pi}\right)^3\left[
n_f \left(\left(\frac{52 \zeta (3)}{9}-\frac{401}{324}\right) C_A
   C_F+\left(\frac{134}{27}-4 \zeta (3)\right)
   C_A^2
\right. \right.\nonumber\\ &&\left.\left.
+\left(\frac{2407}{1458}-\frac{16 \zeta (3)}{9}\right) C_F^2\right)
+n_f^2
   \left(-\frac{697 C_A}{729}-\frac{169 C_F}{1458}\right)
\right]
\ ,
\\
 y_3 & =&   
 \frac{\alpha_s}{4\pi}\left( 
\frac{4 C_F}{3}
\right)
+\left(\frac{\alpha_s}{4\pi}\right)^2\left[
C_F \left(\frac{61 C_A}{27}-\frac{68 n_f}{27}\right)-\frac{4
   C_F^2}{27}
\right]
\nonumber\\&&
+\left(\frac{\alpha_s}{4\pi}\right)^3\left[
n_f \left(\left(\frac{64 \zeta (3)}{9}-\frac{8305}{729}\right)
   C_F^2-\frac{2}{243} (864 \zeta (3)+1079) C_A C_F\right)
\right. \nonumber\\ &&\left.
-\frac{8}{729} (972
   \zeta (3)+143) C_A C_F^2+\left(\frac{32 \zeta (3)}{9}+\frac{6611}{729}\right)
   C_A^2 C_F
\right. \nonumber\\ &&\left.
-\frac{76}{243} C_F n_f^2+\frac{8}{729} (648 \zeta (3)-125) C_F^3
\right]
\ .
\end{eqnarray}
Substitution of these results into (\ref{traceanog}) and (\ref{traceanoq})
leads to the main result of this paper,
\begin{eqnarray}
\lefteqn{\eta_{\mu\nu}\left(T^{\mu\nu}_g\right)_R
%
= \frac{\alpha_s}{4\pi}\left( 
\frac{14}{3} C_F \left(m \bar{\psi }\psi  \right)_R-\frac{11}{6} C_A \left(F^2\right)_R
\right)
+\left(\frac{\alpha_s}{4\pi}\right)^2}
\nonumber  \\&&\;\;\;\;\;\;\;\; \times
\left[
 \left(C_F \left(\frac{812 C_A}{27}-\frac{22 n_f}{27}\right)+\frac{85
   C_F^2}{27}\right) \left(m \bar{\psi }\psi  \right)_R
   + \left(\frac{28 C_A n_f}{27}-\frac{17 C_A^2}{3}+\frac{5 C_F
   n_f}{54}\right)\left(F^2\right)_R
\right]
\nonumber\\&&
+\left(\frac{\alpha_s}{4\pi}\right)^3\left[
 \left\{n_f \left(\left(\frac{368 \zeta
   (3)}{9}-\frac{25229}{729}\right) C_F^2-\frac{2}{243} (4968 \zeta (3)+1423) C_A
   C_F\right)
\right. \right. \nonumber\\ &&\left.\left.
+\left(\frac{32 \zeta (3)}{3}-\frac{91753}{1458}\right) C_A
   C_F^2+\left(\frac{294929}{1458}-\frac{32 \zeta (3)}{9}\right) C_A^2
   C_F-\frac{554}{243} C_F n_f^2
\right.\right. \nonumber\\ &&\left.\left.
+\left(\frac{95041}{729}-\frac{64 \zeta
   (3)}{9}\right) C_F^3\right\}\left(m\bar{\psi } \psi  \right)_R
\right. \nonumber
\\ 
&&\left.
+ \left\{n_f
   \left(\left(\frac{1123}{162}-\frac{52 \zeta (3)}{9}\right) C_A C_F+\left(4
   \zeta (3)+\frac{293}{36}\right) C_A^2+\frac{16}{729} (81 \zeta (3)-98)
   C_F^2\right)
\right. \right.\nonumber
\end{eqnarray}
\begin{eqnarray}
&&\left.\left.
+n_f^2 \left(\frac{655 C_A}{2916}-\frac{361
   C_F}{729}\right)-\frac{2857 C_A^3}{108}\right\}\left(F^2\right)_R
\right]
\  , \label{tgtwoloop}
\\
\lefteqn{\eta_{\mu\nu}\left(T^{\mu\nu}_q\right)_R 
=  
\left(m \bar{\psi }\psi  \right)_R
+ \frac{\alpha_s}{4\pi}\left( 
\frac{4}{3} C_F \left(m \bar{\psi }\psi  \right)_R+\frac{1}{3} n_f
   \left(F^2\right)_R
\right)
+\left(\frac{\alpha_s}{4\pi}\right)^2}
\nonumber\\ 
&& \;\;\;\;\;\;\;\;\;\;\;\;\;\;\;\;   \times
\left[
    \left(C_F \left(\frac{61 C_A}{27}-\frac{68
   n_f}{27}\right)-\frac{4 C_F^2}{27}\right) \left(m \bar{\psi }\psi  \right)_R
   + \left(\frac{17 C_A n_f}{27}+\frac{49 C_F
   n_f}{54}\right)\left(F^2\right)_R
\right]
\nonumber\\&&
+\left(\frac{\alpha_s}{4\pi}\right)^3\left[
 \left\{n_f \left(\left(\frac{64 \zeta
   (3)}{9}-\frac{8305}{729}\right) C_F^2-\frac{2}{243} (864 \zeta (3)+1079) C_A
   C_F\right)
\right.\right. \nonumber\\ &&\left.\left.
-\frac{8}{729} (972 \zeta (3)+143) C_A C_F^2+\left(\frac{32 \zeta
   (3)}{9}+\frac{6611}{729}\right) C_A^2 C_F-\frac{76}{243} C_F
   n_f^2
\right. \right.\nonumber\\ &&\left.\left.
+\frac{8}{729} (648 \zeta (3)-125) C_F^3\right\}\left(m \bar{\psi }\psi  \right)_R
\right. \nonumber\\ &&\left.
+ \left\{n_f
   \left(\left(\frac{52 \zeta (3)}{9}-\frac{401}{324}\right) C_A
   C_F+\left(\frac{134}{27}-4 \zeta (3)\right)
   C_A^2+\left(\frac{2407}{1458}-\frac{16 \zeta (3)}{9}\right) C_F^2\right)
\right.\right. \nonumber\\ &&\left.\left.
+n_f^2
   \left(-\frac{697 C_A}{729}-\frac{169 C_F}{1458}\right)\right\}\left(F^2\right)_R
\right]
\ , \label{tqtwoloop}
\end{eqnarray}
corresponding to the three-loop extension of  (\ref{051}) and (\ref{050}), respectively.
In accord with the general result (\ref{traceanoqgadd2}), we note that
the sum of these two equations reproduces the three-loop expression of (\ref{111}) and is thus
RG-invariant. However,
each of them exhibits the dependence on the RG scale $\mu$, i.e., 
\begin{equation}
T^\lambda_\lambda=\left. \eta_{\lambda\nu}\left(T^{\lambda\nu}_g\right)_R\right|_\mu+
\left. \eta_{\lambda\nu}\left(T^{\lambda\nu}_q\right)_R\right|_\mu\ ,
\label{separatemu}
\end{equation}
due to the contributions of order $\alpha_s^2$ and higher (see the discussion in Sec.~\ref{sec1}).

The formulas (\ref{tgtwoloop})-(\ref{separatemu}) and the other formulas obtained in this section 
should be evaluated using the three-loop running coupling constant, 
which is obtained by solving the RG equation (see (\ref{betaexact}), (\ref{beta1}), (\ref{beta2})),
\begin{equation}
\frac{d\ln \alpha_s}{d\ln \mu^2}=-\beta_0 \frac{\alpha_s}{4\pi}-\beta_1 \left(\frac{\alpha_s}{4\pi}\right)^2
-\beta_2 \left(\frac{\alpha_s}{4\pi}\right)^3\ ,
\label{run}
\end{equation}
as
\begin{eqnarray}
\ln \frac{\mu^2}{\Lambda_{\rm QCD}^2}=
\frac{4
   \pi }{\beta _0 \alpha _s(\mu )}+\frac{\beta _1 }{\beta _0^2}\ln \left(\frac{\beta _0 \alpha _s(\mu
   )}{4 \pi }\right)
+
\frac{\left(\beta _0 \beta _2-\beta _1^2\right) \alpha _s(\mu )}{4 \pi  \beta _0^3}
\ ,
\label{lambdaQCD}
\end{eqnarray}
where the constant of integration is represented by the the QCD scale parameter $\Lambda_{\rm QCD}$ according to the definition in \cite{Chetyrkin:2000yt,Collins:2011zzd}, and  
this result may be further solved for $\alpha _s(\mu )$ iteratively, leading to
\begin{equation}
\frac{\alpha _s(\mu )}{4\pi}=\frac{1}{\beta _0 L}
-\frac{\beta _1 \ln L}{\beta _0^3 L^2}
+\frac{1}{L^3}\left[\frac{\beta _2}{\beta _0^4}+\frac{\beta _1^2 \left(\ln ^2L-\ln L-1\right)}{\beta
   _0^5}\right]
\ ,
 \end{equation}
where $L\equiv \ln \left(\mu^2/\Lambda_{\rm QCD}^2\right)$.
For example, using the RunDec package~\cite{Herren:2017osy},
we obtain,
\begin{equation}
\alpha_s(\mu=1~\rm{GeV})= 0.47358
\ldots\ ,
\label{alphasval}
\end{equation}
as the result of the three-loop evolution in the $\overline{\rm MS}$ scheme,
where, starting from the initial value
$\alpha_s(M_Z)=0.1181$,
the number of active flavors is determined automatically
such that the decoupling is 
performed at the pole mass of the respective heavy quark.
We use this value~(\ref{alphasval}) for numerical evaluations with the $n_f=3$ active flavors.\footnote{The
value (\ref{alphasval}) corresponds to $\Lambda_{\rm QCD}^{(3)}\simeq0.336$~GeV, when (\ref{lambdaQCD}) 
is used 
with $n_f=3$ fixed.}

Before ending this section,
we mention an immediate consequence of our three-loop results,
combined with the exact operator identities in QCD, which read~\cite{Kolesnichenko:1984dj,Braun:2004vf,Tanaka:2018wea} 
\begin{equation}
\partial_\nu T^{\mu\nu}_q = \bar{\psi} gF^{\mu\nu}\gamma_\nu \psi\ ,  
 \label{phys}
\end{equation}
up to the terms which vanish using the equations of motion, $\left(i\Slash{D}-m\right)\psi=0$, and 
\begin{equation}
\partial_\nu T^{\mu\nu}_g = F_{\nu }^{\ \mu} D_\alpha  F^{\alpha\nu}\ . \label{eq2}
\end{equation}
Note that (\ref{phys}) and (\ref{eq2}) are compatible with the condition (\ref{nonren}),
using the equations of motion for the gluon fields, $D_\alpha F^{\alpha\nu}=g\bar{\psi}\gamma^\nu \psi$,
and the fact that the equations of motion are preserved under renormalization.
Then, the $\partial_\mu$-derivative of (\ref{pre}) gives
\begin{eqnarray}
(\bar{\psi}gF^{\mu\nu}\gamma_\nu \psi)_R = (Z_\psi-Z_Q)
\bar{\psi}gF^{\mu\nu}\gamma_\nu \psi
+ Z_K\partial^\mu (m\bar{\psi}\psi)
+\left(Z_B-\frac{Z_Q}{4}\right)\partial^\mu F^2
\ .
\label{rge3b}
\end{eqnarray}
Here, the last term plays roles for cases beyond one loop, 
because $Z_B-\frac{Z_Q}{4}={\cal O}(\alpha_s^2)$ 
from the results (\ref{zqthree}) and (\ref{zbthree}).
It is remarkable that 
(\ref{rge3b}), together with the above results of the three-loop renormalization constants, 
leads to the three-loop evolution equation for the twist-four quark-gluon operator, 
\begin{eqnarray}
\frac{\partial}{\partial \ln \mu}
\left( g\bar\psi F^{\lambda\nu}\gamma_\nu \psi \right)_R
& =&   
 \frac{\alpha_s}{4\pi}\left( 
 \left(-\frac{16 C_F}{3}-\frac{4 n_f}{3}\right)\left( g\bar\psi F^{\lambda\nu}\gamma_\nu \psi \right)_R+\frac{4 
   C_F}{3}\partial^\lambda\left( m \bar\psi \psi \right)_R
\right)
\nonumber\\&&
+\left(\frac{\alpha_s}{4\pi}\right)^2\left[
   \left(\frac{11 C_A}{18}+\frac{4 C_F}{9}\right) n_f \partial^\lambda\left( F^2 \right)_R
   \right.
   \nonumber\\
   &&
   +\left(
   \left(\frac{20 C_F}{9}-\frac{70 C_A}{27}\right)n_f-\frac{752 C_A C_F}{27}+\frac{224
   C_F^2}{27}\right)\left( g\bar\psi F^{\lambda\nu}\gamma_\nu \psi \right)_R 
  \nonumber\\
  &&\left.
    + \left(\frac{122 C_A C_F}{27}-\frac{136 C_F n_f}{27}-\frac{8
   C_F^2}{27}\right)\partial^\lambda\left( m \bar\psi \psi \right)_R
\right]
\nonumber
\end{eqnarray}
\begin{eqnarray}
&&
+\left(\frac{\alpha_s}{4\pi}\right)^3\left[
\partial^\lambda\left( F^2 \right)_R \left(n_f^2 \left(-\frac{56 C_A}{81}-\frac{19 C_F}{27}\right)
+n_f
   \left(\frac{433 C_A C_F}{108}+\frac{1235 C_A^2}{324}+\frac{14
   C_F^2}{27}\right)\right)
   \right. \nonumber\\ &&\left.
   +\left( g\bar\psi F^{\lambda\nu}\gamma_\nu \psi \right)_R \left(n_f \left(\left(16 \zeta
   (3)+\frac{322}{9}\right) C_A C_F+\left(48 \zeta (3)-\frac{3589}{81}\right)
   C_A^2
   \right.\right.\right. \nonumber\\ &&\left.\left.\left.
   +\left(\frac{9878}{243}-64 \zeta (3)\right) C_F^2\right)+n_f^2
   \left(\frac{2116 C_A}{243}+\frac{1358 C_F}{243}\right)+\left(128 \zeta
   (3)+\frac{17056}{243}\right) C_A C_F^2
   \right.\right. \nonumber\\ &&\left.\left.
   -\frac{16}{243} (648 \zeta (3)+2615)
   C_A^2 C_F-\frac{32}{243} (648 \zeta (3)-35) C_F^3\right)
   \right. \nonumber\\ &&\left.
   +\partial^\lambda\left( m \bar\psi \psi \right)_R \left(n_f
   \left(\left(\frac{64 \zeta (3)}{3}-\frac{8305}{243}\right) C_F^2-\frac{2}{81}
   (864 \zeta (3)+1079) C_A C_F\right)
   \right.\right. \nonumber\\ &&\left.\left.
   +\left(-32 \zeta
   (3)-\frac{1144}{243}\right) C_A C_F^2+\left(\frac{32 \zeta
   (3)}{3}+\frac{6611}{243}\right) C_A^2 C_F
    \right.\right. \nonumber\\ &&\left.\left.
   -\frac{76}{81} C_F
   n_f^2+\frac{8}{243} (648 \zeta (3)-125) C_F^3\right)
\right]
\ ,
\label{ee3}
\end{eqnarray}
which extends the previous two-loop result obtained in \cite{Hatta:2018sqd};
the substitution $\left( g\bar\psi F^{\lambda\nu}\gamma_\nu \psi \right)_R
\to - \left(F_{\nu }^{\ \lambda} D_\alpha  F^{\alpha\nu}\right)_R$ in (\ref{ee3})
gives the evolution equation for the twist-four
gluonic operator, $F_{\nu }^{\ \lambda} D_\alpha  F^{\alpha\nu}$.
Those evolution equations should be handled combining with the RG equations for higher twist operators 
arising in the RHS, i.e., (see (\ref{1111}))
\begin{eqnarray}
\frac{d}{d\ln \mu}\ \partial^\lambda \left(F^2\right)_R&=&
2\left(
\beta_0 \frac{\alpha_s}{4\pi}+2\beta_1 \left(\frac{\alpha_s}{4\pi}\right)^{2}+3\beta_2 \left(\frac{\alpha_s}{4\pi}\right)^{3}
\right)
\partial^\lambda\left(F^2\right)_R
\nonumber\\
&-&4\left(\gamma_{m0}\frac{\alpha_s}{4\pi} + 2 \gamma_{m1}\left(\frac{\alpha_s}{4\pi}\right)^2+3\gamma_{m2}\left(\frac{\alpha_s}{4\pi}\right)^3\right)\partial^\lambda\left(m\bar{\psi}\psi\right)_R\ ,
\label{11111}
\\
\frac{d}{d\ln \mu}\ \partial^\lambda\left(m\bar{\psi}\psi\right)_R&=& 0\ ,
\end{eqnarray}
with (\ref{beta0}), (\ref{beta1})-(\ref{gammam2}).

\section{Anomaly-induced mass structure of hadrons}
\label{sec4}

As mentioned in Sec.~\ref{sec1},
it is well-known that the QCD trace anomaly (\ref{111}) is related to the nucleon mass $M$, as
\begin{equation}
2M^2=\langle P|T^\lambda_\lambda|P\rangle
=\langle P| \left( \frac{\beta (g_R)}{2g_R}\left(F^{\lambda\nu}F_{\lambda\nu}\right)_R + \left(1+\gamma_m(g_R)\right) \left(m\bar{\psi} \psi\right)_R \right)|P\rangle\ .
\label{mass}
\end{equation}
This indicates that almost all of the nucleon mass could be attributed to the
quantum loop effects in QCD which induce the trace anomaly.
Based on this equation, it is also argued frequently that, in the chiral limit, the entire mass comes from gluons.
However, the partition of QCD loop effects for the trace anomaly into the gluon and quark contributions, 
(\ref{051}) and (\ref{050}),
and their three-loop extension (\ref{tgtwoloop}) and (\ref{tqtwoloop}),
shows that the latter statement is not correct:
evaluating (\ref{tgtwoloop})-(\ref{separatemu}) with $N_c=3$, $n_f=3$, one finds,
\begin{equation}
2M^2=\langle P|\left. \eta_{\lambda\nu}\left(T^{\lambda\nu}_g\right)_R\right|_\mu|P\rangle+ \langle P|\left. \eta_{\lambda\nu}\left(T^{\lambda\nu}_q\right)_R\right|_\mu|P\rangle\ ,
\label{muindependentsum}
\end{equation}
where
\begin{eqnarray}
\lefteqn{\left. \eta_{\lambda\nu}\left(T^{\lambda\nu}_g\right)_R\right|_\mu= \left(-0.437676
   \alpha _s(\mu)-0.261512 \alpha _s^2(\mu)-0.183827 \alpha _s^3(\mu)\right)\left.\left(F^2\right)_R\right|_\mu}
   \nonumber\\
   &&\;\;\;\;\;\;\;\;\;\;\;\;\;\;\;
   + \left(0.495149 \alpha _s(\mu)+0.776587 \alpha
   _s^2(\mu)+0.865492 \alpha _s^3(\mu)\right) \left(m\bar \psi \psi\right)_R
\ ,
\label{tg3loop1}
   \\
\lefteqn{\left. \eta_{\lambda\nu}\left(T^{\lambda\nu}_q\right)_R\right|_\mu = \left(0.0795775 \alpha _s(\mu)+0.0588695 \alpha _s^2(\mu)+0.0216037 \alpha
   _s^3(\mu)\right)\left. \left(F^2\right)_R\right|_\mu}
   \nonumber\\
 && \;\;\;\;\;\;\;\;\;\;\;\; +
   \left(1+0.141471 \alpha
   _s(\mu)-0.00823495 \alpha _s^2(\mu)-0.0643511 \alpha _s^3(\mu)\right) \left(m\bar \psi \psi\right)_R
\label{tq3loop1}\ ,
\end{eqnarray}
where $\left(m\bar \psi \psi\right)_R$ is independent of the scale $\mu$,
while the $\mu$ dependence of $\left(F^2\right)_R\bigr|_\mu$
is controlled by the RG equation (\ref{1111});
substituting (\ref{alphasval}),  we obtain
\begin{eqnarray}
\left. \eta_{\lambda\nu}\left(T^{\lambda\nu}_g\right)_R\right|_{\mu=1~\rm{GeV}}&=&-0.285452 \left.\left(F^2\right)_R\right|_{\mu=1~\rm{GeV}}+ 0.500593 \left(m\bar \psi \psi\right)_R
\ ,
\label{1gevg}
   \\
\left. \eta_{\lambda\nu}\left(T^{\lambda\nu}_q\right)_R\right|_{\mu=1~\rm{GeV}} &=& 0.0531842 \left.\left(F^2\right)_R\right|_{\mu=1~\rm{GeV}}+1.05832
  \left(m\bar \psi \psi\right)_R
\ .
\label{1gevq}
\end{eqnarray}
Because 
$\alpha^3_s(1~\rm{GeV})\simeq0.1$, the neglected four-loop contributions
are expected to produce corrections less than ten percent. 

For simplicity, in the chiral limit, we find
\begin{equation}
\frac{\langle P|\left. \eta_{\lambda\nu}\left(T^{\lambda\nu}_q\right)_R\right|_\mu|P\rangle}{\langle P| \left. \eta_{\lambda\nu}\left(T^{\lambda\nu}_g\right)_R\right|_\mu|P\rangle}=
-0.181818-0.0258682 \alpha _s(\mu)+ 0.0424613 \alpha _s^2(\mu)\ ,
\end{equation}
so that the gluon- and quark-loop effects make the nucleon mass 
heavy and light, respectively, with the magnitude of the former being five times larger 
than that of the latter, and it appears that the $\mu$-dependence of this result for the relative size 
of the gluon/quark loop effects in the chiral limit is rather weak.
It is also worth noting that the total sum (\ref{muindependentsum}) of (\ref{1gevg}) and (\ref{1gevq}) 
allows us to constrain the matrix element of $F^2$ as
\begin{equation}
\langle P|\left. \left(F^2\right)_R\right|_{\mu=1~\rm{GeV}}|P\rangle\simeq-8.61M^2\ .
\end{equation}
When taking into account the quark-mass effects, 
the matrix element of the quark scalar operator,  $\langle P|m \bar \psi \psi|P\rangle$, participates; 
its value may be constrained 
from the informations on the sigma terms (see, e.g., \cite{Gasser:1982ap,Ji:1994av,Ji:1995sv,Alarcon:2011zs}), but we do not go into the detail here.

Next, we consider the pion case, making the substitution: 
$|P\rangle \to |\pi(p)\rangle$ with $p^2=m_\pi^2$.
Then, (\ref{mass})  becomes
\begin{equation}
2m_\pi^2
=\bigl\langle \pi(p)\bigl| \left( \frac{\beta (g_R)}{2g_R}\left(F^2\right)_R + \left(1+\gamma_m(g_R)\right) \left(m\bar{\psi} \psi\right)_R \right)\bigr|\pi(p)\bigr\rangle\ ,
\label{masspi}
\end{equation}
which implies,
in the chiral limit $m=0$,
\begin{equation}
\left. \bigl\langle \pi(p)\bigl| \left(F^2\right)_R\bigr|\pi(p)\bigr\rangle\right|_{m=0}=0\ .
\end{equation}
Since the PCAC relation ($f_\pi$ is the pion decay constant),
\begin{equation}
- \left(m_u + m_d\right)\langle 0|\left(\bar{u}u +\bar d d \right) |0\rangle = 2f_\pi^2m_\pi^2 \ ,
\end{equation}
indicates $m_\pi^2 \sim m$ as $m\to 0$, 
(\ref{masspi}) gives the relation 
among the ${\cal O}(m)$ terms 
when the substitution $\bigr|\pi(p)\bigr\rangle \to \bigr|\pi(p)\bigr\rangle_0+\bigr|\pi(p)\bigr\rangle_1+\ldots$,
where $\bigr|\pi(p)\bigr\rangle_0\equiv  \left. \bigr|\pi(p)\bigr\rangle\right|_{m=0}$ and $\bigr|\pi(p)\bigr\rangle_1$ 
is the ${\cal O}(m^1)$-term, is made,
such that
$\bigl\langle \pi(p)\bigl|\left(m\bar{\psi} \psi\right)_R \bigr|\pi(p)\bigr\rangle 
\to\  _{0}\bigl\langle \pi(p)\bigl|  \left(m\bar{\psi} \psi\right)_R 
 \bigr|\pi(p)\bigr\rangle_0$
and 
$\bigl\langle \pi(p)\bigl| \left(F^2\right)_R\bigr|\pi(p)\bigr\rangle\to\ _{0}\bigl\langle \pi(p)\bigl|  \left(F^2\right)_R\bigr|\pi(p)\bigr\rangle_1
+\ _{1}\bigl\langle \pi(p)\bigl|  \left(F^2\right)_R\bigr|\pi(p)\bigr\rangle_0$, up to the corrections of ${\cal O}(m^2)$.
On the other hand, the pion mass can also be calculated as the mass shift due to the ordinary
first-order perturbation theory in the quark mass term in the QCD Hamiltonian,
as~\cite{Gasser:1982ap}
\begin{equation}
m_\pi^2=\ _{0}\bigl\langle \pi(p)\bigl| m\bar{\psi} \psi
 \bigr|\pi(p)\bigr\rangle_0\ ,
\label{chiralp}
\end{equation}
and, combining this with (\ref{masspi}), we obtain
\begin{eqnarray}
\left(1-\gamma_m(g_R)\right)m_\pi^2
&=&\bigl\langle \pi(p)\bigl|  \frac{\beta (g_R)}{2g_R}\left(F^2\right)_R\bigr|\pi(p)\bigr\rangle
\ ,
\label{masspi3}
\end{eqnarray}
to ${\cal O}(m)$ accuracy. Therefore, up to the corrections of ${\cal O}(m^2)$, the terms associated with the $F^2$ operator and the $m\bar \psi \psi$ operator  in the RHS
of  (\ref{masspi})
contribute to $m_\pi^2$ according to the relative weights,  $\left(1-\gamma_m(g_R)\right)$ and
$\left(1+\gamma_m(g_R)\right)$, respectively, where 
\begin{equation}
\gamma_m(g_R)= 0.63662 \alpha _s+ 0.768352 \alpha _s^2+ 0.801141 \alpha _s^3\simeq 0.559\ ,
\end{equation}
using (\ref{gammam0})-(\ref{gammam2}), and (\ref{alphasval}).
Substituting (\ref{chiralp}) and (\ref{masspi3}) into
(\ref{muindependentsum})
with (\ref{tg3loop1})-(\ref{1gevq}) and $M^2 \to m_\pi^2$,
we find
\begin{eqnarray}
\lefteqn{\frac{1}{2m_\pi^2}\bigl\langle \pi(p)\bigl|\left. \eta_{\lambda\nu}\left(T^{\lambda\nu}_g\right)_R\right|_\mu\bigr|\pi(p)\bigr\rangle}
\nonumber\\
&&\;\;\;\;\;\;\;\;
=0.611111-0.12215 \alpha
   _s(\mu)-0.124659 \alpha _s^2(\mu)
-0.0430357 \alpha _s^3(\mu)\ ,
\\
\lefteqn{\frac{1}{2m_\pi^2}\bigl\langle \pi(p)\bigl|\left. 
\eta_{\lambda\nu}\left(T^{\lambda\nu}_q\right)_R\right|_\mu\bigr|\pi(p)\bigr\rangle}
\nonumber\\
&&\;\;\;\;\;\;\;\;
=0.388889+0.12215 \alpha _s(\mu)+0.124659 \alpha _s^2(\mu)
+0.0430357 \alpha _s^3(\mu)
 \ ,
\end{eqnarray}
and
\begin{eqnarray}
&&
\frac{1}{2m_\pi^2}\bigl\langle \pi(p)\bigl|\left. \eta_{\lambda\nu}\left(T^{\lambda\nu}_g\right)_R\right|_{\mu=1~\rm{GeV}}\bigr|\pi(p)\bigr\rangle=0.521\ ,
\\
&&
\frac{1}{2m_\pi^2}\bigl\langle \pi(p)\bigl|\left. 
\eta_{\lambda\nu}\left(T^{\lambda\nu}_q\right)_R\right|_{\mu=1~\rm{GeV}}\bigr|\pi(p)\bigr\rangle=0.479
 \ ,
\end{eqnarray}
which hold to ${\cal O}(m)$ accuracy. As in the nucleon case, the $\mu$-dependence of the result is rather weak,
but this result shows the structure which is completely different from the nucleon case:
both the gluon- and quark-loop effects make the pion mass 
heavy, each of them giving rise to roughly half of the pion mass.
This remarkable feature may be eventually attributed to particular nature of the pion as a Nambu-Goldstone
boson, but the clarification needs further studies, including those to make connections with the QCD analysis of hadron masses
based on the decomposition of the QCD Hamiltonian~\cite{Ji:1994av,Ji:1995sv,Lorce:2017xzd} into the quark/gluon components.
 
\section{Conclusions}
\label{sec5}

We have studied the renormalization of the energy-momentum tensor in QCD at three-loop order
and derived the analytic formulas for all the relevant renormalization constants in the $\overline{\rm MS}$ scheme.
To achieve this, we did not need any new loop calculation.
We explained our method which allows us to determine the structure of the renormalization mixing involving the twist-four as well as twist-two
operators, from the knowledge of the basic RG functions, $\beta(g)$ and $\gamma_m(g)$,  and the anomalous dimensions for twist-two, spin-2 operators,
which are available to the required accuracy in the literature.
This method utilized, in particular, the constraints from the fact that the trace of the total energy-momentum tensor is determined completely by the basic RG functions in the RG-invariant form, and that the renormalization constants in the MS-like schemes
obey the divergent pole structures independent of any mass.

Determining the renormalization structure at three loops, we were immediately 
able to calculate the trace contributions for  the quark and gluon
parts of the energy-momentum tensor separately, leading to the decomposition of 
the three-loop QCD trace anomaly into the quark and gluon parts. 
Our result extends the previous result in the two-loop order into three loops, and 
the features found in the former case are preserved at the three-loop level,
although their analytic expressions become considerably complicated and cumbersome.
Going from the two-loop formula to the three-loop formula, 
we have found that the accuracy is generally improved from ten percent level to a few percent level, and, therefore,
our three-loop formula should provide good control on the uncertainties in the calculations for most practical purposes.
Our formula obtained in the $\overline{\rm MS}$ scheme will be most convenient for various actual applications,
but it would be interesting to derive the similar formula using other regularization schemes,
e.g., the gradient flow regularization \cite{Suzuki:2013gza,Makino:2014taa},
and study the scheme dependence.

As an immediate application, we used our three-loop formula to analyze the anomaly-induced mass
structure of nucleon as well as of pion. Our three-loop formula allowed us to carry out such analysis 
at a few percent-level accuracy,
and provided a new insight on the hadron masses such that they are generated from the quantum anomaly effects
for quarks as well as for gluons. Our analysis revealed that the nature of the masses, in particular, 
the dominant roles played by the quark part and the gluon part of the energy-momentum tensor, are completely different between nucleon and pion. 
Another application, which was not discussed in this paper,
is
to constrain the gravitational form factor $\bar{C}_{q,g}$ of a hadron, as treated in \cite{Hatta:2018sqd}. 
$\bar{C}_{q,g}$ 
receives much attention 
in connection with the force distribution inside the nucleon~\cite{Polyakov:2018zvc,Polyakov:2018exb,Hatta:2018ina,Lorce:2017xzd,Lorce:2018egm}
and the nucleon's transverse spin sum rule~\cite{Hatta:2012jm,Leader:2012ar,Chakrabarti:2015lba}.
Quantitative analysis for the constraints on $\bar{C}_{q,g}$
will be discussed elsewhere.

\section*{Acknowledgments}
The author thanks Yoshitaka Hatta and Abha Rajan for stimulating discussions in the collaboration for \cite{Hatta:2018sqd} 
which triggered this work.  

\bibliography{../bibdata/bibdatabase}
\end{document}